# Evidence of large magnetic cooling power and double glass transition in $Tb_5Pd_2$

Mohit K. Sharma and K. Mukherjee*

School of Basic Sciences, Indian institute of Technology, Mandi, Himachal Pradesh -175005, India

*E-mail: kaustav@iitmandi.ac.in

**Abstract**

We report a detailed dc magnetization, ac susceptibility and magnetocaloric properties of a binary intermetallic compound $Tb_5Pd_2$. Our dc magnetization and heat capacity results reveal absence of long range ordering in this compound. Two distinct frequency dependent peaks ($T_{f1}$~60 K and $T_{f2}$~21 K) had been observed in the ac-susceptibility. Analysis of these frequency dependent peaks by Mydosh parameter, power and Vogel-Fulcher law reveals that around $T_{f1}$ the compound is at the boundary of spin glass (SG) like-cluster glass (CG) like state and it undergoes a cluster glass-like freezing below $T_{f2}$. Zero field cooled memory effect and non-linear dc susceptibility also confirmed the presence of two glassy transitions in this compound. The transformation from SG/CG boundary to CG phase was also confirmed by the magnetic relaxation measurement and Arrott plots study. Remarkably, a significant magnetic entropy change was also observed in the temperature range of 60-120 K. Additionally a large relative cooling power also observed in this compound. The observed value is comparable to those of promising refrigerant material in this temperature range and is quite notable as in this compound magnetic hysteresis was absent in this temperature range. It was noted that in this compound short-range interactions persist up to a higher temperature above $T_{f1}$ and this is responsible for the observation of significant MCE over a wide temperature range. Our studies suggest that this compound is an example of a glassy magnetic compound which shows large magnetocaloric effect.

# I. INTRODUCTION

Investigation of the magnetic materials exhibiting significant magnetocaloric effect (MCE) has been a subject of extensive research in past couple of decades due to their interesting properties both from fundamental and application points of view [1-4]. Magnetic refrigeration (MR) is based on the MCE and has attracted attention because of highly efficient and environment friendly cooling in comparison to conventional gas compression/expansion methods [4]. The materials which exhibit the large MCE at low-temperature region can be suitable for application in space science, liquefaction industry, however, for domestic and industrial refrigeration, the large MCE is required near to room temperature [3]. Literature report indicates that systems which show first order phase transition (FOPT) usually have a large value of MCE. However, these materials have some drawbacks like high thermal and magnetic hysteresis, thereby, reducing the refrigeration efficiency [5, 6]. Materials showing a paramagnetic to ferromagnetic transition also show significant MCE. Theoretically, based on magnetic frustration, new type of materials showing large MCE has been proposed [7, 8]. Experimental literature reports suggest that such materials showing large MCE in the frustrated glass-like magnetic state are relatively rare [9-11]. Hence, at present, investigation in this area is being focussed on materials which show optimal magnetocaloric properties like, large magnetic entropy change ($\Delta S_M$), high relative cooling power (RCP) in the operating temperature range along with minimal magnetic hysteresis. Apart for technological applications, from the viewpoint of basic physics, investigation of these materials is interesting as one can investigate the complex magnetic phases, nature of magnetic phase transitions and phase coexistence which are present in these materials [11-16].

Many intermetallic compounds formed by blending rare earth (R) and other nonmagnetic/magnetic elements have been found to show good MCE. In this context, a family of binary intermetallics $R_5Pd_2$ (R= rare-earth ions) are being investigated in recent years. Members of this family are known to be good magnetocaloric materials and show a complex magnetic state [17-23]. Klimczak et al., reported that the magnetic properties of $R_5Pd_2$ compounds [17]. Large $\Delta S_M$ is reported in the compounds $Ho_5Pd_2$ and $Dy_5Pd_2$ [18, 20-21]. Complex magnetic behaviour and presence of glassy magnetic state was reported for $Er_5Pd_2$ [22]. For $Tb_5Pd_2$ compound a complex cluster glass state has been reported [19].

However, a detailed study about the nature of magnetic transition and along with the investigation of MCE of this compound is lacking in literature.

In the present paper, we report a detailed study magnetic and magnetocaloric properties of $Tb_5Pd_2$. Results of dc magnetization and heat capacity divulge absence of long range ordering. It was observed from ac susceptibility studies that in this compound there are two frequency dependent peaks ($T_{f1}$~60 K and $T_{f2}$~21). These peaks were analysed by Mydosh parameter, power and Vogel-Fulcher law. Our results reveal that around $T_{f1}$ the compound is at the boundary of spin glass like-cluster glass like state and it undergoes a cluster glass-like freezing below $T_{f2}$. This observation of double glassy transition was also confirmed from zero field cooled memory effect and non-linear dc susceptibility measurements. Magnetic relaxation measurements and Arrott plots study also reveal the transformation from spin glass/cluster glass boundary to cluster glass phase. Notably a significant MCE and large relative cooling power was also observed in the temperature range of 60-120. In this compound short-range interaction persist up to a higher temperature above $T_{f1}$ and this is responsible for the observation of significant MCE over a wide temperature range. The observed values are comparable to those of promising refrigerant material in this temperature range. Additionally it was noted magnetic hysteresis was absent in this temperature range, satisfying another important criteria for a magnetic refrigerant material. Thus our study reveals that this compound is an example of a glassy magnetic compound which shows large magnetocaloric effect.

## II. EXPERIMENTAL DETAILS

Polycrystalline $Tb_5Pd_2$ had been synthesized by arc melting techniques using a stoichiometric ratio of Tb and Pd (>99.99 % purity) under the similar conditions as reported in [18]. X-ray diffraction (XRD) was performed at room temperature in the range 20°-70° (steps size $0.02^0$) using Rigaku diffraction. Figure 1 shows the room temperature indexed X-ray diffraction (XRD) pattern for the compound. The compound crystallizes in cubic structure and the obtained lattice parameter was 13.597 Å which was in accordance with literature report [17]. In order to get confirmation about the stoichiometry of the compound, we performed energy-dispersive x-ray spectroscopy. The average atomic stoichiometry was found to be in accordance with the expected values. Magnetic field and temperature dependent, DC and AC magnetization measurements were performed using Magnetic Properties Measurements System while the temperature dependent heat capacity measurement was performed using Physical Properties Measurements System, both from Quantum Design, USA.

## III. RESULTS AND DISCUSSION

### A. DC Magnetization study

Fig. 2 shows the temperature response of zero field cooled (ZFC) dc magnetic susceptibility ($\chi = M/H$) recorded under various applied fields (100 Oe to 10 kOe). For the lowest field i.e. 100 Oe, the curve shows a sharp peak around 60 K, indicating the presence of a magnetic phase transition. Inset (I) of Fig. 2 displays the inverse magnetic susceptibility curve at 100 Oe fitted with Curie-Weiss (CW) law in the range of 100 to 300 K. The effective magnetic moment ($\mu_{eff}$) and CW temperature ($\theta_p$) obtained from fitting were found to be around 10.7 $\mu_B$/Tb atom and 75 K respectively. The calculated effective magnetic moment was near the theoretical value (9.72 $\mu_B$/Tb atom) while, the positive value of $\theta_p$ indicates the dominance of ferromagnetic interactions. As noted from the inset (II) of Fig. 2(a), the bifurcation between the ZFC and FC curves starts from ~ 85 K, which is defined as the irreversibility temperature ($T_{irr}$). The enormous separation between ZFC and FC curves revealed the presence of considerable magnetic anisotropy and/or glassy magnetic phase in the compound. With the increase in applied field it is observed that the sharp peak is significantly broadened. Additionally, a field-induced transition was observed around 40 K at 5 kOe. On further increase of magnetic field this transition becomes more prominent, while the transition temperature decreases. Hence the observed broadening might arise due to the development of a field-induced ordered state or some due to some additional glass-like magnetic phase transition.

### B. Magnetic field response of magnetization

In order to shed light on the nature of the low temperature magnetic phase of the compound, magnetic field response of magnetization was carried out in the temperature range of 10 to 150 K. Magnetic hysteresis was observed at low temperatures, however, this feature was absent above 40 K. Figure 3 shows some representative isotherms of the compound at selected temperatures. Below the 60 K, the curves are symmetric and non-saturating (up to ±70 kOe field) which reveals the random behavior of magnetic moments and indicates the presence of glassy dynamics in the compound [23 24]. To substantiate the above statement, coercive field ($H_C$) and magnetic retentivity ($M_R$) (calculated by the magnetic isotherms data) are plotted as a function of temperature (I and II inset of Fig. 3). Both the curves are fitted with exponential function of the form [25 26]:

$$H_C(T) = H_C(0)\exp(-\alpha_1 T)$$

$$M_R(T) = M_R(0)\exp(-\alpha_2 T) \ldots\ldots\ldots(1)$$

where $H_C(0)$ and $M_R(0)$ are the coercivity and retentivity at 0 K, respectively, $\alpha_1$ and $\alpha_2$ are the constants. The values of $\alpha_1$ and $\alpha_2$ obtained from the fitting are respectively 0.144±0.002 and 0.128±0.002 while that of $H_C(0)$ and $M_R(0)$ are 24.26±0.02 kOe and 12.01±0.1 kOe respectively. Different systems show different type behavior is observed for temperature response of $H_C$ and $M_R$. For example, for a single domain particle, $H_C$ varies as $T^{1/2}$ [27] while in the weak antiferromagnetic system it shows parabolic behavior [28]. The feature observed in our case is analogous to that observed for a glass-like magnetic state [25 26]. Additionally, temperature response of heat capacity ($C$) at zero field does not show any sharp peak (III inset of Fig. 3), which indicates of absence the long range ordering in the compound and is in accordance to the observation of glass-like magnetic state in this compound.

**C. Evidences of the presence of double glass transition**

Although ac susceptibility behaviour of this compound had been reported in literature, for the sake of completeness, we discuss the temperature dependence of ac susceptibility. Fig. 4 (a and b) represents the temperature response of real ($\chi'$) and imaginary ($\chi''$) part of ac susceptibility, measured at different frequencies. The $\chi'$ curves show a sharp peak around 60 K ($T_{f1}$) which shifts towards higher temperature with increasing the frequency (Fig. 4(a)). Such shifting in peak temperature is a characteristic of spin glass (SG) and/or cluster glass (CG) magnetic state [19, 29]. This peak temperature coincides to that observed in dc magnetization. Additionally, $\chi'$ it shows a weak hump like feature in the low temperature region (Fig. 4 (a)). A more vivid manifestation of the temperature response of frequency dependence of the compound is exhibited in the imaginary part. $\chi''$ being the Fourier transform of the two spin-correlation function, characterizes the dynamics of magnetic system and displays anomalous behavior near magnetic phase transition. It is noticed that the $\chi''$ curve (Fig. 4(b)), shows one additional frequency dependent peak around 21 K ($T_{f2}$), along with the peak at 60 K. The characterization of both these peaks was done through Mydosh parameter [29]

$$\delta T_f = [\Delta T_f/(T_f(\Delta \log f))] \ldots\ldots(2)$$

and power law [29]

$$\tau = \tau_0 (T_f/T_g - 1)^{-z\nu} \ldots\ldots(3)$$

where $T_g$ is the true glass transition temperature, $T_f$ is the freezing temperature, $\tau_0$ is the relaxation time of single spin flip and $z\nu$ is the exponent. The scaling of $\tau$ with the reduced temperature $\varepsilon = [(T_f/T_g - 1)]$ was represented in the insets I and II of Fig. 4. The parameters obtained from the equations (2) and (3) were tabulated in Table 1. Generally for a SG system $\delta T_f$ varies from 0.005 to 0.01 and $\tau_0$ is near $10^{-13}$ sec, whereas for a CG glass system $\delta T_f$ varies from 0.02 to 0.06 and a much slower relaxation rate is observed [29, 30]. The difference in relaxation time values around both the freezing temperatures explicitly suggests a significant change in the size and distribution of clusters around these temperatures. The increment in relaxation time with decreasing the temperature is arises due to the increasing average cluster size. The values of the extracted parameters for $T_{f1}$ suggest a behaviour that is at the boundary of SG like and CG like state, while, that for $T_{f2}$ suggest a CG like state. This type to double glass-like state in a compound is not uncommon and has been reported in literature [31].

Vogel-Fulcher (V-F) law [29] can be used to differentiate between different types of glass freezing. V-F law is mathematically expressed as

$$\tau = \tau_0 \exp (E_a / k_B (T_f - T_0)) \ldots\ldots (4)$$

where $\tau_0$ is the characteristic relaxation time, $k_B$ is the Boltzmann constant, $T_0$ is the VF temperature measuring the intercluster interaction strength and $E_a$ is the activation energy separating different metastable state. From the parameters obtained from this fitting, it is noted that $E_a/k_B T_0 \sim 0.6$ for $T_{f1}$ and $E_a/k_B T_0 \sim 10$ for $T_{f2}$. This parameter $E_a/k_B T_0$ is quite large for a CG system while for spin-glasses it is less than 3 [31, 32]. Hence the result of this analysis also supports the conclusions of the previous paragraph.

To give further evidences of the presence of double glass transition in this compound ZFC memory effect measurement by using the stop and wait protocol was done [30]. ZFC memory effect is a unique signature of arising from cooperative spin-spin freezing phenomenon and is not observed in superparamagnetic systems, thereby, helpful in discerning the glassy state arising out of freezing and blocking phenomenon [33, 34]. In this measurement while cooling the compound in zero field two halts was made for 2 h at 50 and 20 K (shown in Fig. 4(c)). The reference curve (ZFC$_{ref}$) (the normal ZFC curve obtained after cooling to the lowest temperature without any halt) was subtracted from the corresponding heating curve (ZFC$_{tw}$) performed after the halts. The difference between these two curves is shown in the inset of Fig. 4(c). The results show two clear dips around both the stop temperature, affirming the existence of memory effect, thereby, confirming the presence glass-like freezing in the compound.

To further crosscheck the presence of double glass transition investigation of non-linear dc susceptibility was carried out, as it is powerful technique which gives the details of local magnetic behavior and shows some anomaly near the glass transition temperatures. The non-linear parts of dc susceptibilities are obtained from temperature and field dependence of magnetization. The non-linear susceptibilities were expressed as, $M = \chi_1 H + \chi_3 H^3/3! + \chi_5 H^5/5!$; where $\chi_1$ is the linear magnetic susceptibility, and the higher order terms of $\chi$ are the non-linear magnetic susceptibility. The data for this measurement was carried out using the protocol described in Ref [22, 35]. Fig. 4 (d) represents the temperature response of the third order susceptibility ($\chi_3$). It was noted that $\chi_3$ is negative throughout the range of measurements and show two peaks around $T_{f1}$ and $T_{f2}$. Hence our observation of ZFC memory effect and non-linear dc susceptibility unequivocally affirm the presence of double glass-like behaviour in the compound.

**D. Temperature dependent magnetic relaxation study**

In order to further understand the dynamics of glassy magnetic state, the magnetic relaxation measurement was carried out by using the protocol: the compound was cooled in zero field from a temperature in the paramagnetic region to the measuring temperatures. Once the measurement temperature was reached, a magnetic field of 50 Oe was applied for 20 min. After that the field was switched off and magnetization was noted as function of time. The magnetization vs. time plot for various temperatures (10 to 100 K) was shown in Fig. 5. Among the various functional forms that have been proposed to describe the change of magnetization with time, the curves were best fitted with the following expression [36]

$$M = M_0 - S \ln(1+t/t_0) \quad \ldots\ldots (5)$$

where $M_0$ is initial magnetization measured at a constant field and $S$ is the magnetic viscosity. Good fits are obtained for all the curves and the fitting parameters are summarized in Table 2. It is to be noted that a logarithmic relaxation implies a distribution of energy barrier in the system. The value of $M_0$ is consistent with FC magnetization behavior of the compound. The value of $S$ was very less in the paramagnetic region and it increases around $T_{f1}$, indication the presence of weak clustering effect. Below $T_{f1}$, the value of S increases as the size of the clusters grows from small spin clusters. Below 40 K, the value of S decreases which might be attributed to enhancement of inter-cluster frustration, leading to CG-like freezing below $T_{f2}$. Additionally, the magnetic relaxation was analyzed by using an interacting magnetic particles model (not shown) [37];

$$W(t) = \frac{d \ln M(t)}{dt} = At^b \dots\dots\dots\dots\dots (6)$$

where *b* is the exponent and it depends on the concentration or strength of magnetic interactions and *A* is the factor related to temperature. The parameter *b* was found to be nearly constant around the $T_{f1}$, which indicates the signature of SG/CG boundary while it was found temperature dependent around the $T_{f2}$, indicating CG behavior [31]. Hence, relaxation measurement also reaffirms the SG/CG boundary to CG transformation in this compound. Hence the results of the magnetic relaxation measurements also support the conclusion of previous sections.

**E. Arrott plots study**

In order to have an idea about the nature of transitions (around $T_{f1}$ and $T_{f2}$) in this compound, the Arrotts plot study was carried out. Arrott plots are an indispensable tool to identify the order of a transition. According to Banerjee's criterion, this Arrott plot exhibits a negative and positive slope for first and second order nature of magnetic transition Ref [38]. Using virgin *M* (*H*) isotherms, *H/M* curves were plotted as a function of $M^2$, as shown in Fig. 6. As noted from the figure, a positive slope was observed around 60 K, while, a negative slope was observed below 40 K. This observation indicates that the transition around $T_{f1}$ is second order in nature while the one around $T_{f2}$ is first order in nature. From the figure it was observed that there was a change in curvature below 40 K which is possible associated to increasing the cluster size. These curvatures further increase with decreasing temperature which related to CG freezing [39]. The increasing behavior of $H_C$ and $M_R$ with decreasing temperature (as shown in insets of Fig. 3) also support the size increment nature of the clusters. To confirm the exact temperature (where first order transition is started), we analyzed the Arrott plot data in terms of magnetic free energy $F(M, T)$. Landau's expansion of free energy, can be expressed as [40]

$$F(M,T) = \frac{C_1}{2}M^2 + \frac{C_3}{4}M^4 + \frac{C_5}{6}M^6 + \dots\dots\dots - HM \dots\dots\dots\dots(7)$$

where $C_1(T)$, $C_3(T)$ and $C_5(T)$ are the Landau coefficients. These coefficients are calculated by using the equation [11, 14]

$$H/M = C_1(T) + C_3(T) M^2 + C_5(T) M^4 \quad \ldots\ldots\ldots (8)$$

The sign of coefficient $C_3$ determines the order of phase transition; for the negative value the transition is first order, while, it is a second order for the positive value of $C_3$. The calculated value of $C_3$ was plotted as function of temperature (not shown). The curve positive value above 40 K and it changes its sign below this temperature, which was in accordance to our observation from Arrott plots. From the above measurement results it was noticed that there are two type of transition present in this compound.

**F. Magnetocaloric effect**

We further investigate the compound by the temperature dependent magnetocaloric effect (MCE) in terms of the isothermal magnetic entropy change ($\Delta S_M$). The $\Delta S_M$ was derived from the virgin curve of $M(H)$ isotherms (in the temperature range of 30-170 K). For this, each isotherm was measured after cooling the respective compound from room temperature to the measured temperature and $\Delta S_M$ is calculated using the following expression [41]

$$\Delta S_M = \Sigma \left[ (M_n - M_{n+1})/(T_{n+1} - T_n) \right] \Delta H_n \quad (9)$$

where $M_n$ and $M_{n+1}$ are the magnetization values measured at field $H_n$ and $H_{n+1}$ at temperature $T_n$ and $T_{n+1}$ respectively. Selected temperature dependent $\Delta S_M$ curve for the different applied field was shown in Fig. 7 (a). The results show $Tb_5Pd_2$ have a significant $\Delta S_M$ value over a wide temperature range of 60 to 120 K with maximum value of ~ 7.3 J/kg-K, for 70 kOe around 80 K. The observed value of $\Delta S_M$ was significant in this temperature range [42-44]. Moreover, in this compound magnetic hysteresis was absent in this temperature range, thereby, satisfying another important criteria for a magnetic refrigerant material. Additionally, it was observed that this curve was asymmetric around the peak value which is in accordance to the absence of long range ordering in this compound, as, symmetric temperature evolution of $\Delta S_M$, was generally observed in compounds having long-ranged magnetic ordering [2]. Additionally, the asymmetric behavior in $\Delta S_M$ was also reported for systems having spin fluctuations and/or spin flop effects [2, 11 and 45]. It was reported from theoretical predictions that in the paramagnetic region the $\Delta S_M$ varies as $H^2/2T^2$. Upper inset of Fig. 7 (b) shows the fitting at selected temperatures. It was noted that, the true paramagnetic region comes above the 100 K, with our magnetization results. Hence it can be said that in this compound short-range interactions persist up to a higher temperature above $T_{f1}$, this is responsible for the observation of significant MCE over a wide temperature range.

In order to check the practical utility of any material which can be used as magnetic refrigerant, the relative cooling power (RCP) is calculated. The RCP is the measure of the amount of heat transfer between cold and hot reservoir in an ideal refrigeration cycle. The RCP is defined as the product of maximum $\Delta S_M$ ($\Delta S_M^{Max}$) and full width of half maximum of the peak in $\Delta S_M$ ($\Delta T_{FWHM}$), i.e.

$$\text{RCP} = \Delta S_M^{max} \times \Delta T_{FWHM} \quad \ldots\ldots (10)$$

However, in our sample, the $\Delta S_M$ has distributed asymmetrically. Therefore, we calculated the RCP by using the equation [46 47];

$$RCP = -\int_{T_h}^{T_c} |\Delta S_M(T, \Delta H)| dT \quad (11)$$

where $T_c$ and $T_h$ are the temperatures of cold and hot ends respectively. The temperature dependent RCP value at different field change was showed in Fig. 7 (b). The observed RCP values are comparable to other reported good magnetic refrigerant in this temperature range [42, 48-50]. It was also noted that both $\Delta S_M$ and RCP values does not show any sign of saturation even at a field of 70 kOe, implying that a enhanced performance can be anticipated at higher applied fields. Furthermore, a crossover was noted below 40 K in $\Delta S_M$ curve, which may be arise due to the existence first order transition and is consistent with the Arrott plots and Landau parameter analysis.

## IV. CONCLUSION

In summary, a detailed investigation of the magnetic and magnetocaloric properties of $Tb_5Pd_2$ was carried out. Analysis of ac susceptibility, magnetic memory effect and nonlinear dc-susceptibility results confirmed the presence of double glass-like behaviour in the compound. Around $T_{f1}$ the compound is at the boundary of SG like and CG like state and it transforms to a CG state below $T_{f2}$. The alteration from SG/CG boundary to CG phase is confirmed by the magnetic relaxation measurement and Arrott plots studied. Moreover, a significant MCE and RCP value also detected in the temperature range 60 to 120 K, where no magnetic hysteresis is observed. Thus our study gave evidences that $Tb_5Pd_2$ belong to that rare category of glassy compound which show large MCE.

">**ACKNOWLEDGEMENTS**

The authors acknowledge the experimental facilities of Advanced Materials Research Center (AMRC), IIT Mandi. Financial support from the SERB project EMR/2016/00682 is also acknowledged.

Table 1. Parameter obtained from Mydosh Parameter and power law fitting

|  | $\delta T_f$ | $T_g$ (K) | $zv$ | $\tau_0$ (sec) |
|---|---|---|---|---|
| $T_{f1}$ | 0.011 | 59 | 4.5±0.3 | ~ $10^{-10}$ |
| $T_{f2}$ | 0.053 | 17 | 7.9±0.2 | ~ $10^{-7}$ |

Table 2. Parameter obtained from logarithmic fitting of magnetization vs. time curves

| $T$ (K) | $M_0$ (emu/gm) | S (emu/gm) |
|---|---|---|
| 10 | 0.0106±0.0004 | 0.00108±0.00003 |
| 20 | 0.0173±0.0006 | 0.00211±0.00006 |
| 40 | 0.0197±0.0002 | 0.00219±0.00003 |
| 60 | 0.0070±0.0003 | 0.00025±0.00001 |
| 80 | 0.0039±0.0003 | 0.00011±0.00001 |
| 100 | 0.0026±0.0002 | 0.00004±0.000002 |

**List of figure**

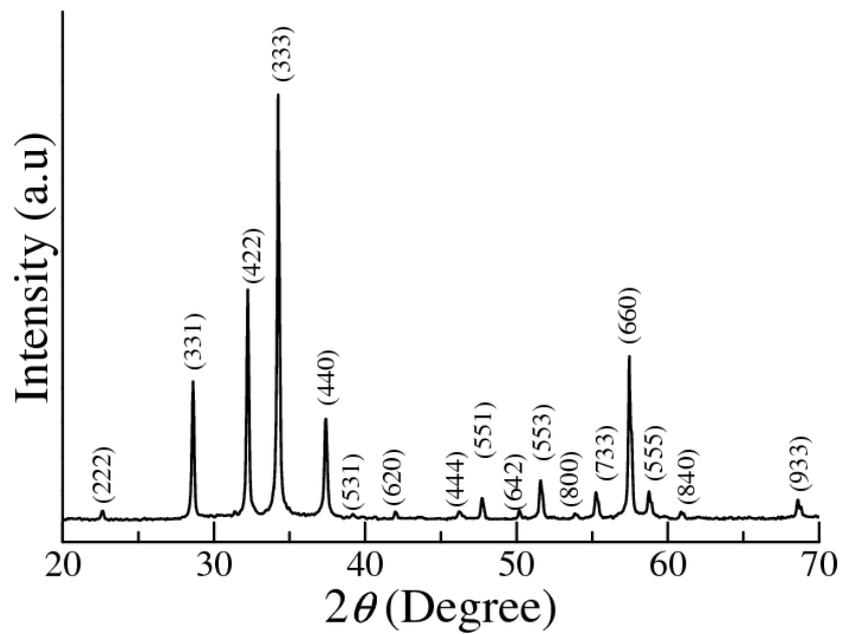

FIG. 1 Room temperature indexed XRD pattern of $Tb_5Pd_2$

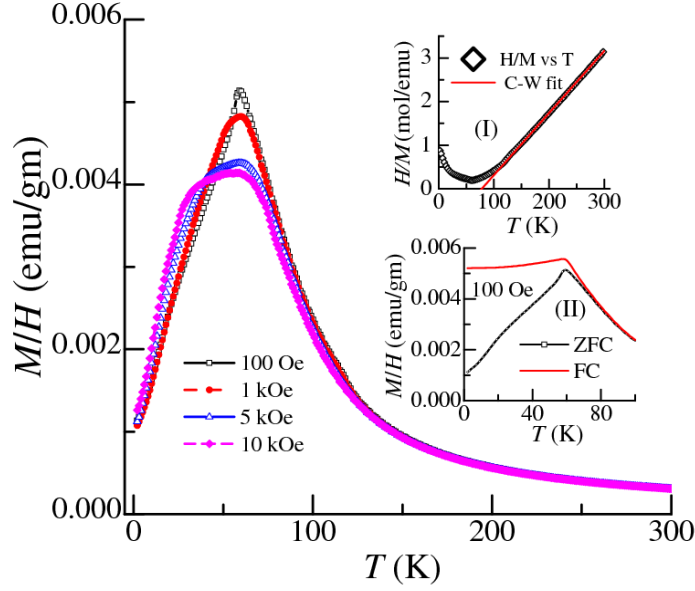

FIG. 2. Temperature response of zero field cooled dc magnetic susceptibility curves measured under various applied field (100 Oe to 10 kOe). Inset (I): Curie Weiss fit of $\chi^{-1}$ vs $T$ at 100 Oe in the temperature range 100-300 K. Inset (II): Temperature response of zero field cooled and field cooled dc magnetic susceptibility at 100 Oe.

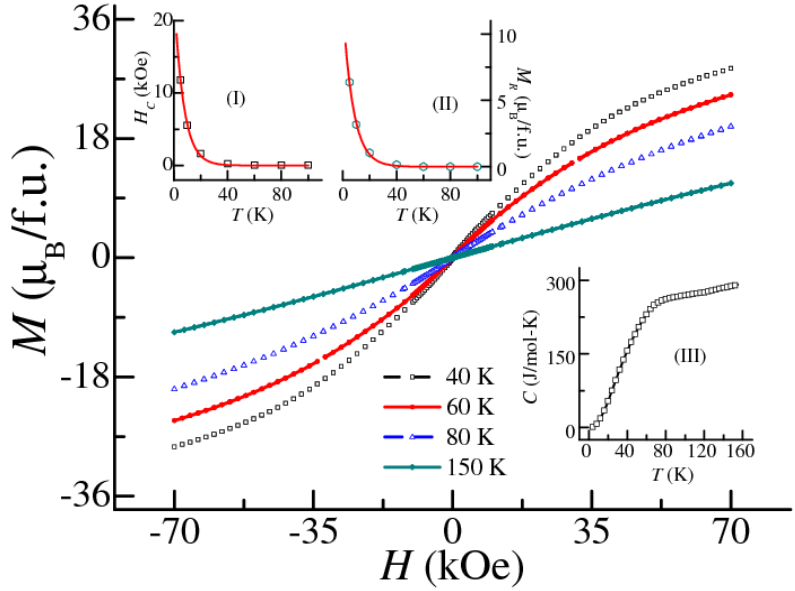

FIG. 3. Magnetic isotherms $M(H)$ at different temperatures. Inset I: Coercive field vs. temperature plot fitted with the equation $H_C(T) = H_C(0) \exp(-\alpha_2 T)$ (red solid line). Inset II: Magnetic retentivity vs. temperature plot fitted with the equation $M_R(T) = M_R(0) \exp(-\alpha_1 T)$ (red solid line). Inset III: Temperature response of heat capacity in zero field.

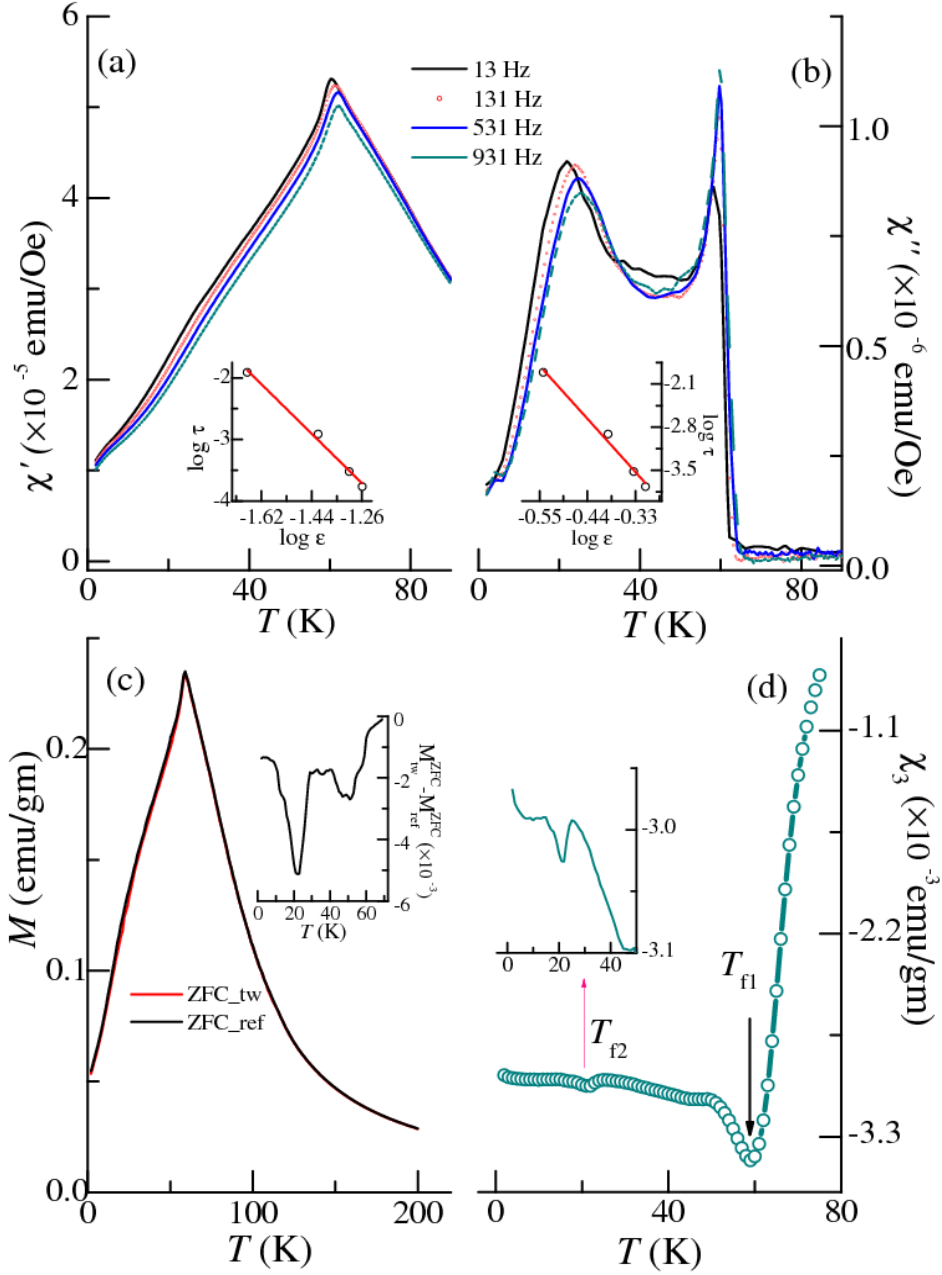

FIG. 4. (a) Temperature response of real part of AC susceptibility ($\chi'$) measured at different frequencies. Inset Power law fit (around $T_{f1}$) of relaxation time ($\tau$) as a function of reduced temperature ($\varepsilon$) using equation (3). (b) Temperature response of imaginary part of AC susceptibility ($\chi''$) at different frequencies. Inset: Power law fit (around $T_{f2}$) of relaxation time ($\tau$) as a function of reduced temperature ($\varepsilon$) using equation (3). (c) Temperature dependent memory effect of $Tb_5Pd_2$ at different waiting temperatures under ZFC condition. Inset: Temperature response of ($M^{ZFC}_{tw} - M^{ZFC}_{ref}$) in the range 2-60 K. (d) Temperature response of third order dc susceptibility ($\chi_3$). Inset: Zoomed out region near $T_{f2}$.

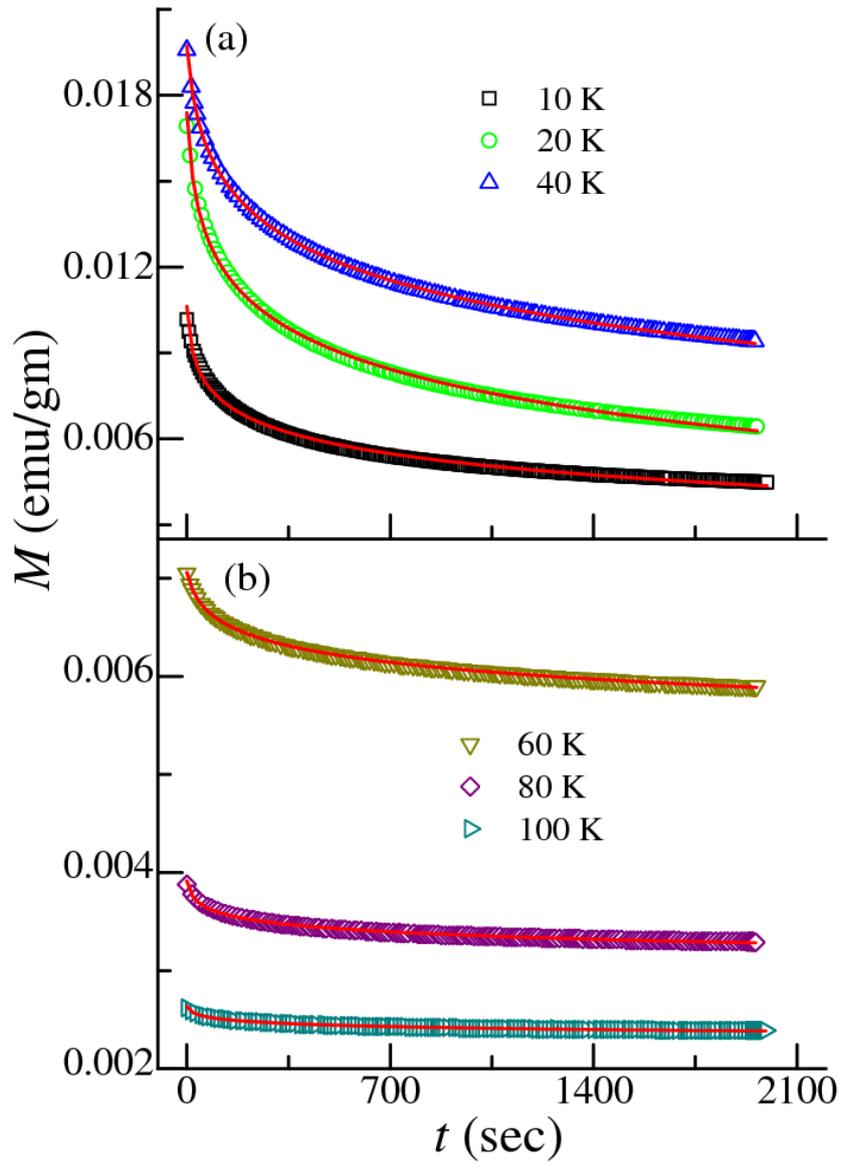

FIG. 5. (a) and (b) Magnetization as a function of time at different temperatures. Solid red line represents logarithmic fit using equation (5).

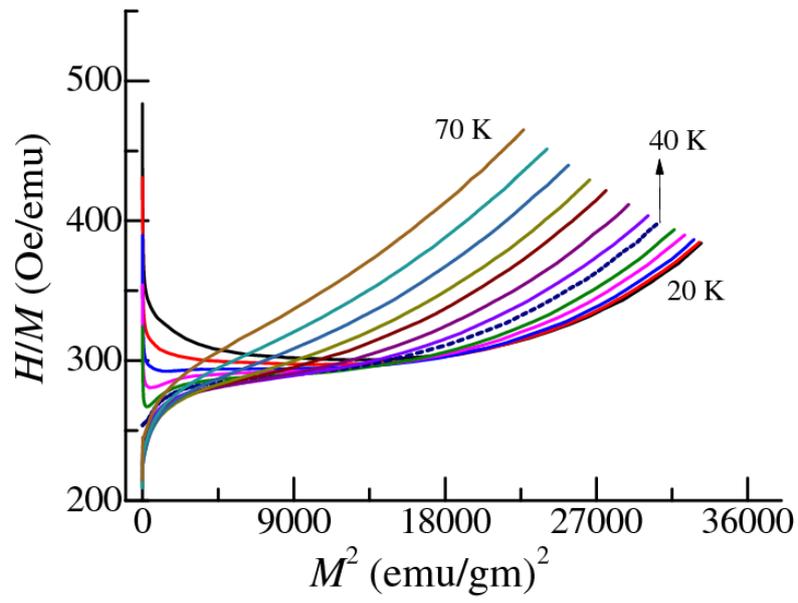

FIG. 6. Arrott plots ($H/M$ vs. $M^2$) for $Tb_5Pd_2$ in the temperature range of 20 to 70 K.

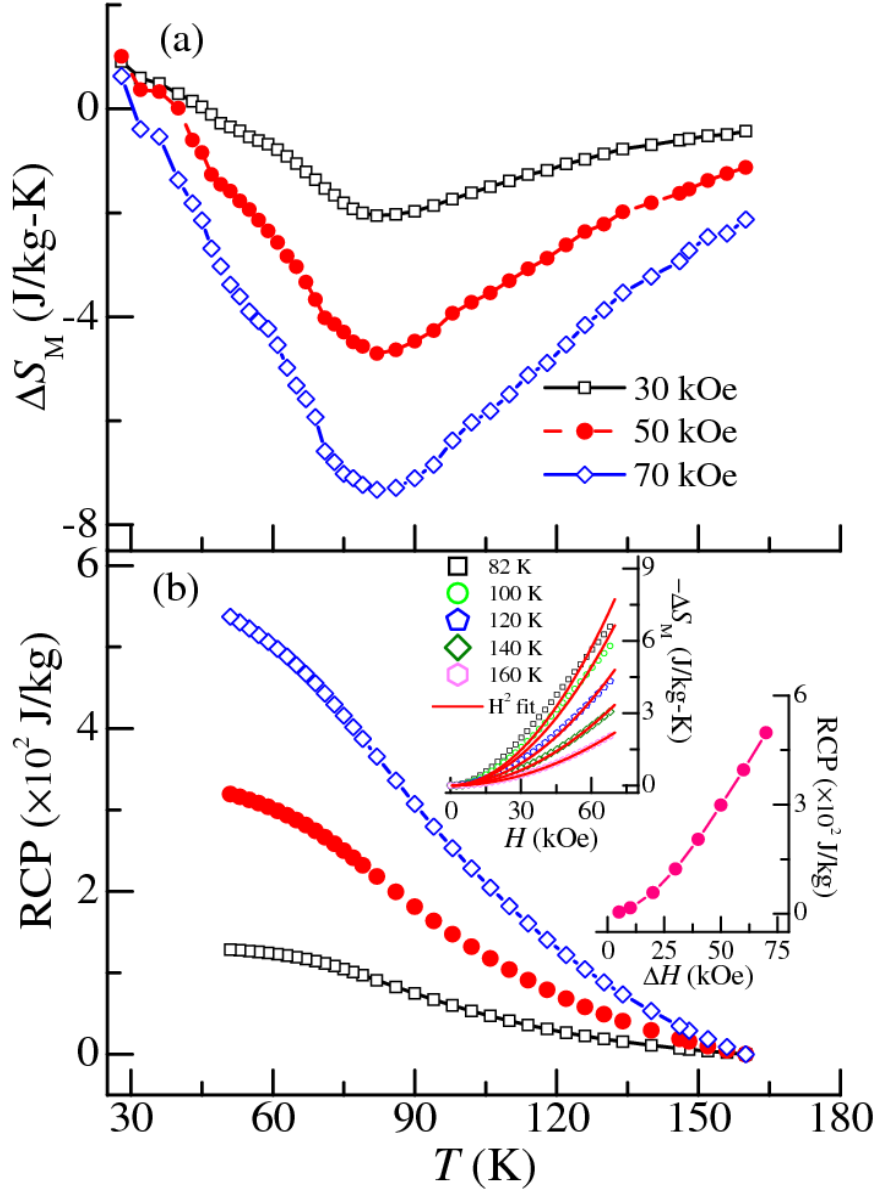

FIG 7 (a) Temperature dependent magnetic entropy change ($\Delta S_M$) under different applied magnetic field change. (b) RCP as function of temperature at different applied field change. Upper inset: $H^2$ fitting for $\Delta S_M$ vs. $H$ curve. Lower inset: Relative cooling power (RCP) plotted as function of $\Delta H$.